\documentclass{emulateapj}
\usepackage{lscape}
\usepackage{longtable}

\newcommand{\noprint}[1]{}

\begin{document}

\title{Characterizing the Cool Kepler Objects of Interest. New Effective Temperatures, Metallicities, Masses and Radii of Low-Mass Kepler Planet-Candidate Host Stars}

\author{Philip S. Muirhead\altaffilmark{1,2}, Katherine Hamren\altaffilmark{3}, Everett Schlawin, B{\'a}rbara~Rojas-Ayala\altaffilmark{4}, Kevin R. Covey\altaffilmark{5}, James P. Lloyd}
\affil{Department of Astronomy, Cornell University, 122 Sciences Drive, Ithaca, NY 14853, USA}

\altaffiltext{1}{Current address: Department of Astronomy, California Institute of Technology, 1200 East California Boulevard, MC 249-17, Pasadena, CA  91125, USA}

\altaffiltext{2}{{\tt philm@astro.caltech.edu}}

\altaffiltext{3}{Current address: Department of Astronomy and Astrophysics, University of California Santa Cruz, 1156 High Street, Santa Cruz, CA  95064, USA}

\altaffiltext{4}{Current address: Department of Astrophysics, American Museum of Natural History, Central Park West at 79th Street, New York, NY 10024, USA.}

\altaffiltext{5}{Hubble Fellow}

\date{}

\begin{abstract}

We report stellar parameters for late-K and M-type planet-candidate host stars announced by the Kepler Mission.  We obtained medium-resolution, K-band spectra of 84 cool ($T_{\rm eff} \lesssim 4400$ $K$) Kepler Objects of Interest (KOIs) from Borucki et al.  We identified one object as a giant (KOI 977); for the remaining dwarfs, we measured effective temperatures ($T_{\rm eff}$) and metallicities [M/H] using the $K$-band spectral indices of Rojas-Ayala et al.  We determine the masses ($M_\star$) and radii ($R_\star$) of the cool KOIs by interpolation onto the Dartmouth evolutionary isochrones.  The resultant stellar radii are significantly less than the values reported in the Kepler Input Catalogue and, by construction, correlate better with $T_{\rm eff}$. Applying the published KOI transit parameters to our stellar radius measurements, we report new physical radii for the planet candidates.  Recalculating the equilibrium temperatures of the planet-candidates assuming Earth's albedo and re-radiation fraction, we find that 3 of the planet-candidates are terrestrial-sized with orbital semi-major axes that lie within the habitable zones of their host stars (KOI 463.01, KOI 812.03 and KOI 854.01).  The stellar parameters presented in this letter serve as a resource for prioritization of future follow-up efforts to validate and characterize the cool KOI planet candidates.

\end{abstract}


\maketitle

\section{Introduction}

Estimating physical parameters of stars which host exoplanets is crucial for estimating the physical parameters of the exoplanets themselves.  The wealth of detailed observations of the Sun has enabled precise calibration of stellar evolution models for Sun-like stars, such that the determination of fundamental stellar physical parameters (mass, effective temperature, luminosity, radius) from observed colors and spectra is routine and generally robust \citep[e.g.][]{Kurucz1991, Nordstrom2004, Valenti2005}.  

For cool dwarfs ($T_{\rm eff} \lesssim 4400 \, K$, $M \lesssim 0.5 \, M_\Sun$), however, the situation is more complex.  Low-mass stellar models are not as well calibrated, and their predictions differ substantially depending on assumptions such as the mixing length parameter.  There are few M dwarfs that are bright enough and nearby enough for direct accurate parallax and radius measurements \citep[e.g.][]{Segransan2003,Berger2006}.  Eclipsing binaries have been the primary source of radii for M dwarfs, but there is a discrepancy between observed radii and predictions from stellar evolution models \citep{Ribas2006,Torres:2011uq}.  The rapid rotation of these close binary systems may be responsible for the discrepancy; however magneto-hydrodynamic effects may supress convection in their interiors, so these radii may not be representative of field objects \citep{Chabrier:2007fj,Kraus:2011kx}.  

Recently, M dwarfs have received increased attention in both transit and radial velocity searches for exoplanets \citep[e.g.][]{Charbonneau2009, Bean2010a, Johnson2010, Mahadevan2010, Muirhead2011, Bonfils2011}, and exoplanet characterization efforts \citep[e.g.][]{Bean2010b, Desert2011, Croll2011}, thanks to the higher detectability and characterization signals from orbiting low-mass exoplanets \citep{Nutzman2008}.  In February of 2011 the Kepler Mission announced 997 objects whose light curves are consistent with the presence of transiting planets \citep[][]{Borucki2011}, 74 of these Kepler Objects of Interest (KOIs) have $T_{\rm eff} < 4400$ $K$ in the Kepler Input Catalog \citep[KIC;][]{Batalha2010, Brown2011}.  A statistical analysis of the KOIs by \citet{Howard2011} reveals a substantial rise in the frequency of short-period, 2-4 $\rm R_\Earth$ planets with decreasing $T_{\rm eff}$ of their host stars, implying that the low-mass planet candidates detected around low-mass stars represent a ubiquitous population of planets in the Galaxy.

Stellar parameters in the KIC were inferred from a photometric survey of stars in the Kepler field-of-view.  However, \citet{Brown2011} state that the KIC stellar parameters are reliable for Sun-like stars, but are ``untrustworthy'' for stars with $T_{\rm eff}$ less than 3750 $K$.  The most reliable estiamtes of M dwarf masses and radii are derived by combining measured stellar luminosities with reliable mass-luminosity relations \citep[e.g.][]{Delfosse2000} and mass-radius relations predicted by stellar evolutionary models \citep[e.g.][]{Baraffe1998}, often with corrections to account for discrepancies between measured and predicted radii \citep[e.g.][]{Torres2007}.  Unfortunately, the low-mass KOIs do not have parallax measurements, which are necessary to estimate stellar luminosity and hence mass and radius with these methods.

Near-infrared spectroscopy offers a more robust method for determining physical parameters for low-mass stars that lack parallax measurements.  The $K$-band (2.0 to 2.4 $\mu m$) contains several useful spectral diagnostics including continuum regions sensitive to $T_{\rm eff}$ \citep{Covey2010} and absorption features that are sensitive to stellar metallicity \citep{Rojas2010, Rojas2012}, with minimal sensitivity to interstellar reddening.

In this letter we report $T_{\rm eff}$ and [M/H] measurements of 84 low-mass KOIs using $K$-band spectroscopy.  We interpolated $T_{\rm eff}$ and [M/H] onto the Dartmouth evolutionary isochrones \citep{Dotter2008, Feiden2011}, which reproduce measurements from optical long-baseline interferometry (OLBI) relatively well (see Figure \ref{fig:models}) and contain a large spread of metallicity grid-points, as required for reliable interpolation of stellar parameters.  We report interpolated stellar masses and radii of the low-mass KOIs, and recalculate the planetary parameters based on the transit parameters in \citet{Borucki2011}.

\section{Observational Classification}\label{obs}

\subsection{Observations}

Observations were carried out with the TripleSpec Spectrograph at the Palomar Observatory 200-inch Hale Telescope.  TripleSpec is a near-infrared slit-spectrograph covering 1.0 to 2.5 $\mu m$ simultaneously with a resolution of 2700 \citep{Herter2008}.  Two positions on the slit, A and B, were used for each target, and exposures were taken in an ABBA pattern with 60 second integration times at each position.  Multiple ABBA sets were taken and combined until each spectrum had a median per-channel signal-to-noise of at least 60.  

For telluric calibration we used SIMBAD to identify a grid of A0V stars in the Kepler field-of-view, and developed an observing sequence such that each KOI observation has a corresponding A0V star observation taken within 40 minutes and with an airmass difference less than 0.1.

\subsection{Target Selection}

We observed all of the KOIs with KIC-ascribed effective temperatures less than 4400 $K$ over 7 nights in June of 2011.  Of the 74 KOIs, 4 appeared to be double objects in the TripleSpec slit viewer with separations of less than 6 arcseconds (roughly the size of the Kepler Spacecraft's point spread function): KOI 326, KOI 641, KOI 249 and KOI 51.  KOI 667 consisted of 5 objects within a 6 arcsecond radius.  These objects are not included in this survey. 

In August of 2011 we observed 15 additional KOIs with KIC-ascribed $T_{\rm eff}$ greater than 4400 $K$, but with colors indicative of low-mass stars: $(J-K)>0.7$, $(r-i)>0.3$, or $(g-r)> 1.0$.  We include 13 of these KOIs in this letter, as their spectra revealed CO features indicative of low-mass stars.  In total, we obtained spectra of 82 KOIs.

\begin{figure}[h!]
\begin{center}
\includegraphics[width=6.5in,angle=90]{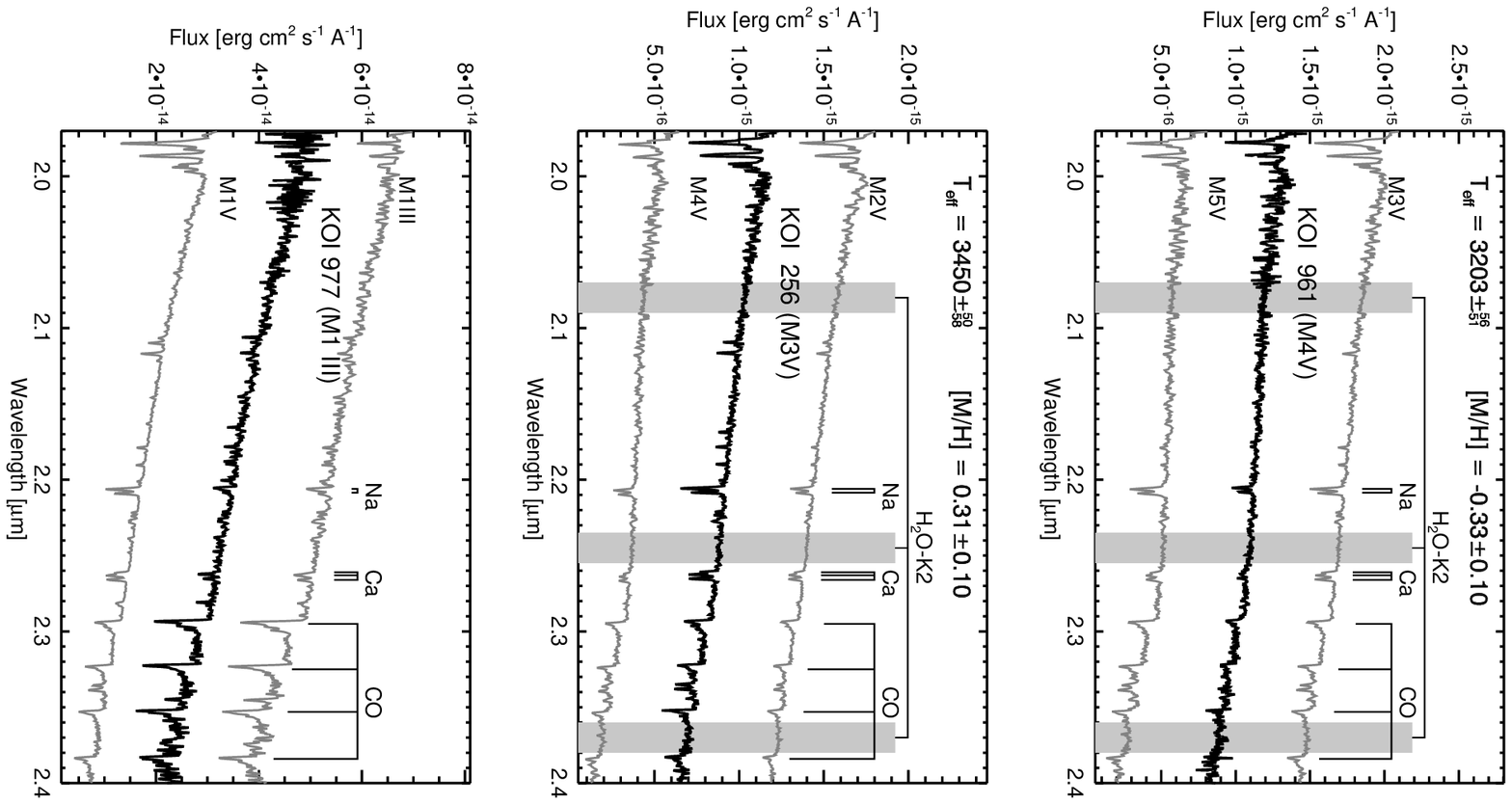}
\caption{\scriptsize Palomar-TripleSpec $K$-band spectra of cool KOIs (black) with comparison KHM spectral type standards from the IRTF Spectral Library \citep[gray,][]{Cushing2005, Rayner2009}. The templates are adjusted to the same scale as the KOI spectra using a ratio of the median flux in K band, and then artificially offset.  We used the $\rm H_2O$-K2 index to compute $T_{\rm eff}$, which is calculated using regions dominated by water opacity, and we used the equivalent widths of the Na I doublet and Ca I triplet to measure [M/H] \citep{Rojas2012}.  {\it Top}: KOI 961 is an example of a metal-poor star with little Na I and Ca I absorption \citep{Muirhead2012}.  {\it Middle}: KOI 256 is an example of a metal-rich star with deep Na I and Ca I absorption.  The metallicities [M/H] are included, with uncertainties accounting for both random and systematic errors.  {\it Bottom}: KOI 977 has a spectrum indicative of a giant with deep CO features but relatively weak Na I and Ca I absorption.}
\label{spectra_plot}
\end{center}
\end{figure}

\subsection{Data Reduction}\label{data_analysis}

The spectra were extracted using a version of the Spextool program modified for the Palomar TripleSpec Spectrograph \citep[][; M. Cushing, private communication 2011]{Cushing2004}. Spextool accepts data in ABBA format.  The {\tt xtellcor} package within Spextool accepts spectra of A0 stars and compares them to a model spectrum of Vega to identify and remove telluric absorption lines in a target spectrum \citep{Vacca2003}.  Figure \ref{spectra_plot} plots three example spectra, with templates of similar spectra type and relevant spectral features indicated.  The templates are taken from the IRTF Spectral Library \citep{Cushing2005, Rayner2009}.

One star in our sample has a K-band spectrum consistent with a giant star, suggesting that the observed light curve is due to a stellar, rather than planetary, companion, or that the transit signal is due to an unresolved blend with an eclipsing binary.  KOI 977 shows weak Na I and Ca I absorption and strong CO absorption, which qualitatively match IRTF template spectra of giant stars but not dwarf stars of the same spectral type.  The spectrum is included in Figure \ref{spectra_plot}, with a giant and dwarf template for comparison.  

\section{Measurement of $T_{\rm eff}$ and [M/H]}

To measure $T_{\rm eff}$ and [M/H] of the remaining dwarfs, we measured three spectral indices from the K-band spectra: the equivalent widths of the Na I and Ca I lines, at 2.210 and 2.260 $\rm \mu m$ respectively, and an index describing the change in flux between three 0.02 $\rm \mu m$-wide bands dominated by water opacity--centered at 2.245, 2.370, 2.080 $\rm \mu m$--called the $\rm H_2O$-K2 index.  \citet{Rojas2012} describe the measurement of the Na I and Ca I equivalent widths, introduce $\rm H_2O$-K2 index, and derive relations between the spectral indices and $T_{\rm eff}$, overall metallicity ([M/H]) and KHM spectral type.  

Briefly, we note that the \citet{Rojas2012} [M/H] relation was calibrated empirically using nearby M dwarfs with F, G or K-type binary companions that have SPOCS [M/H] measurements \citep{Valenti2005}.  Metallicity measurements for stars earlier than M0 ($T_{\rm eff} \gtrsim 4000$ K) represent an extrapolation of the M dwarf [M/H] calibration.  The $T_{\rm eff}$ is calculated by interpolating the [M/H] measurement and $\rm H_2O$-K2 index onto a theoretical surface of $T_{\rm eff}$ versus [M/H] and $\rm H_2O$-K2 index calculated to the BT-settl late-type model spectra of \citet{Allard2011}.  To validate the $T_{\rm eff}$ measurement method, we compare $T_{\rm eff}$ measurements by \citet{Rojas2012} to measurements from optical long-baseline interferometry in Figure \ref{comparison}, Panel A.

The $T_{\rm eff}$ surface is very metallicity insensitive ($<10 $K offsets due to metallicity effects) for $3200<T_{\rm eff}< 3900$, and [M/H]$<0.3$.  For higher temperatures the $\rm H_2O$-K2 index saturates, where the saturation $T_{\rm eff}$ depends on [M/H].  For stars with $\rm H_2O$-K2 near the saturation, a slightly higher $\rm H_2O$-K2 index converts to a large increase in the measured $T_{\rm eff}$.  We accommodate this by providing asymmetric uncertainty estimates in $T_{\rm eff}$ using a Monte Carlo approach described in Section \ref{errors}.  KOI 904 and KOI 956 had $\rm H_2O$-K2 outside of the calculated surface, and are therefore not included in our results.  	
\section{Determination of Mass and Radius}\label{models}

\begin{figure}[]
\begin{center}
\includegraphics[width=3.5in]{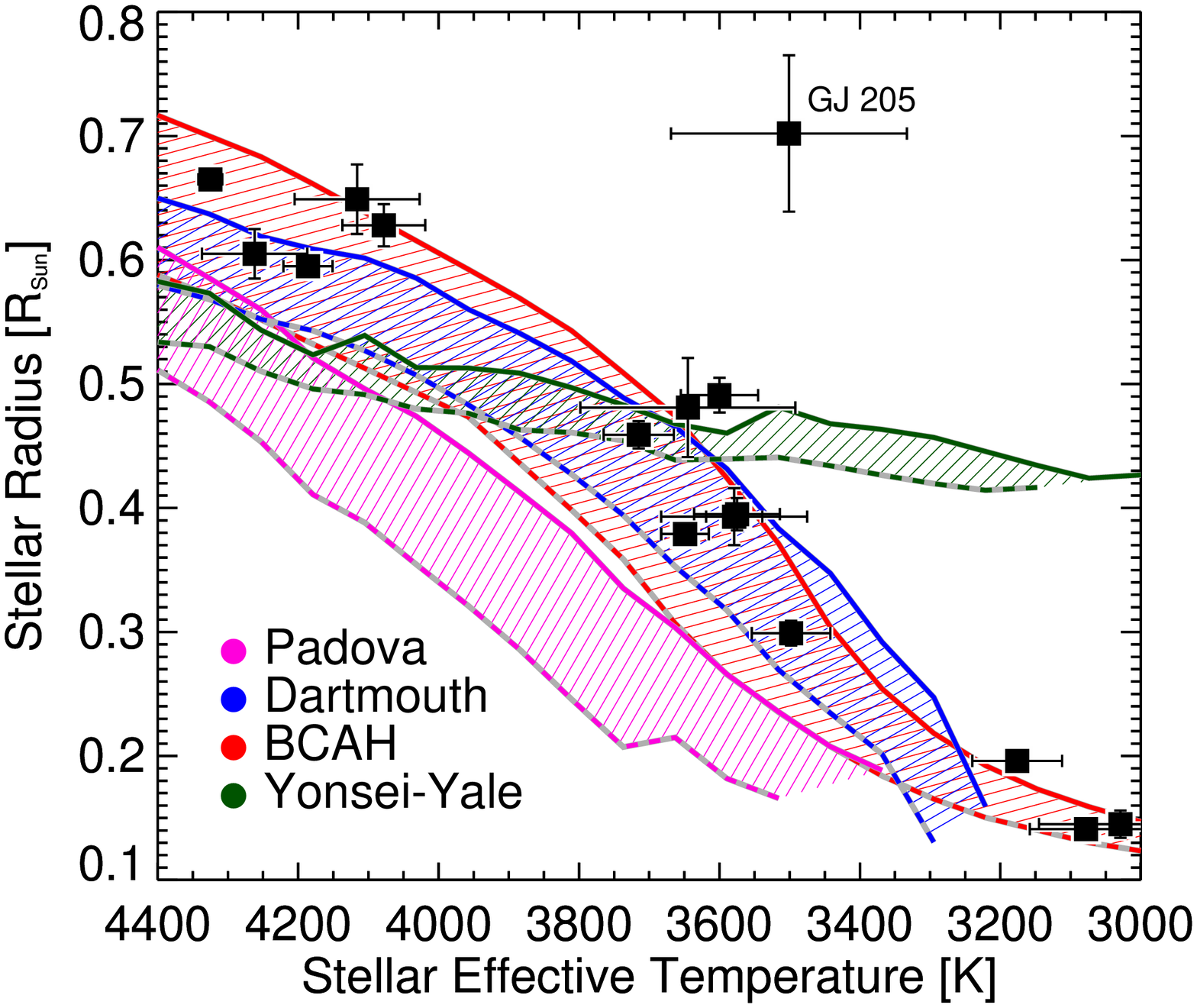}
\caption{Predicted $R_\star$ versus $T_{\rm eff}$ for the 5-Gyr isochrones of Padova \citep{Girardi:2002lr}, Dartmouth \citep[e.g.][]{Dotter2008, Feiden2011}, Yonsei-Yale \citep{Yi:2003fu},  and BCAH \citep{Baraffe1998}, for metallicities between [M/H] = -0.5 ({\it dashed lines}) and [M/H] = 0.0 ({\it solid lines}).  We include empirical measurements of $R_\star$ and $T_{\rm eff}$ for low-mass field stars using optical long-baseline interferometry (OLBI), with values taken from the literature \citep{Berger2006, vanbelle2009, Boyajian2008, Lane2001, Demory2009, Segransan2003, Kervella2008, vonbraun2011}.  The Yonsei-Yale and Padova isochrones do not match the observations as well as the BCAH and Dartmouth isochrones.  The BCAH isochrones are only available in 2 metallicities: [M/H] = 0.0 and [M/H] = -0.5, necessitating poorly constrained interpolation and extrapolation if used to determine stellar mass and radius.  Therefore, we chose to interpolate our $T_{\rm eff}$ and [M/H] measurements of the cool KOIs onto the Dartmouth isochrones.  The OLBI measurements for GJ 205 \citep{Segransan2003} do not match any isochrone predictions, and this is likely the result of systematic errors in the OLBI measurements.  We note that in the latest release of {\it Kepler} planet-candidates, \citet{Batalha2012} confined the stellar parameters of the host stars to the Yonsei-Yale isochrones.  This will produce systematically larger radii for the low-mass stars than is evidenced by long-baseline interferometry.}
\label{fig:models} 
\end{center}
\end{figure}

We place the low-mass KOIs on a grid of physical parameters based on the Dartmouth stellar evolution models \citep{Dotter2008, Feiden2011}.  These models are in generally good agreement with OLBI observations (see Figure \ref{fig:models}), but there may well be systematic offsets in mass, radius or effective temperature, so these inferred physical parameters should be used with caution.  We do not use the BCAH evolution models, as they are only available in two metallicities, [M/H] = 0.0 and [M/H] = -0.5, and a comprehensive metallicity grid is required for reliable interpolation of our measurements.

\begin{figure*}[]
\begin{center}
\includegraphics[width=7.0in]{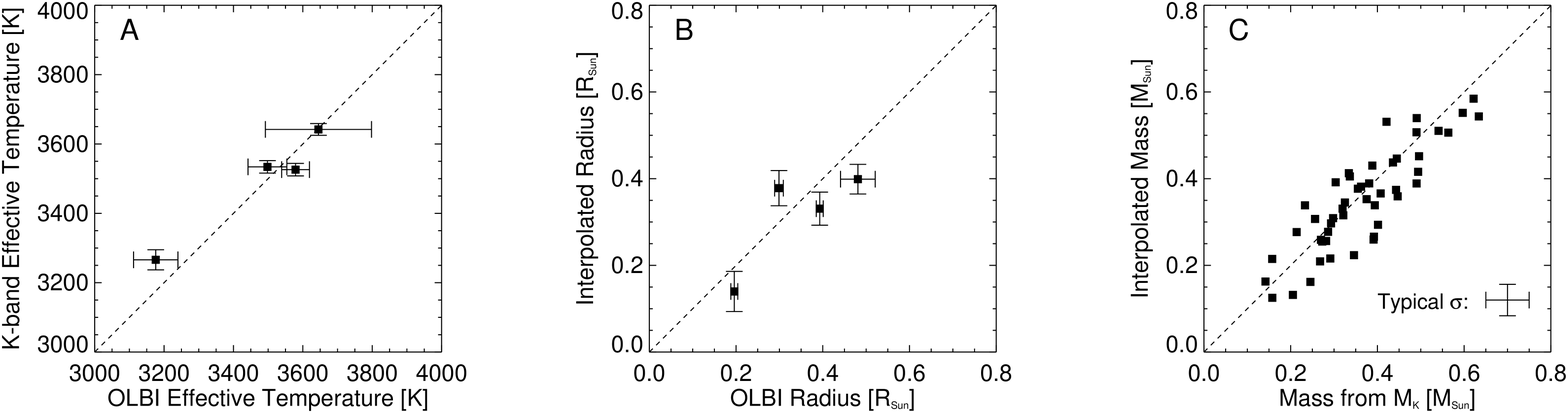}
\caption{Comparison between $T_{\rm eff}$, $R_\star$, and $M_\star$ as measured by optical long-baseline interferometry (OLBI) or the \citet{Delfosse2000} mass-$M_K$ relation ({\it x axes}), and the methods used in this paper ({\it y axes}) for nearby stars with overlapping measurements.  Only 5 stars in \citet{Rojas2012} have empirical $T_{\rm eff}$ and $R_\star$ measurements in the literature: GJ 699, GJ 411 \citep[both by][]{Lane2001}, GJ 525 \citep{Berger2006}, GJ 581 \citep{vonbraun2011} and GJ 205 \citep{Segransan2003}.  GJ 205 is excluded owing to its aberrant OLBI measurements (see Figure \ref{fig:models}).  There are 106 stars in \citet{Rojas2012} with parallaxes in the literature and 2MASS $K$-band measurements, allowing for the calculation of $M_K$ and use of the empirically calibrated \citet{Delfosse2000} mass-$M_K$ relation.  However, we only include those stars with $K$-band effective temperatures that are greater than 3200 K (spectral types of M4 and earlier), and we only include stars with parallax measurements larger than 50 $mas$, amounting to 56 stars.  The methods for measuring low-mass star masses and radii employed in this letter are consistent with empirical measurements to within the calculated uncertainties.}
\label{comparison} 
\end{center}
\end{figure*}

\begin{figure}[htbp]
\begin{center}
\includegraphics[width=3.5in]{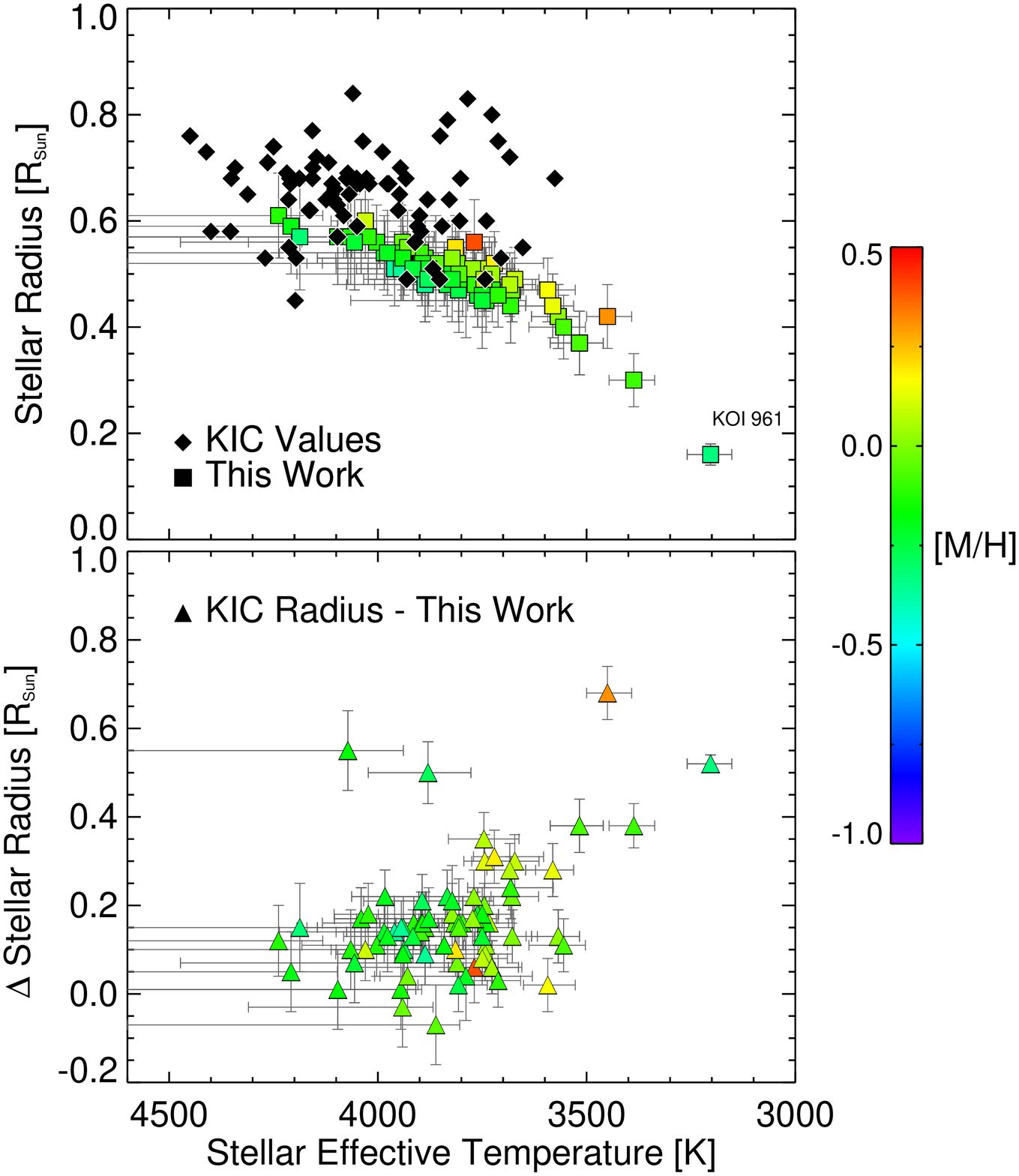}
\caption{Temperature, metallicity and radius determinations for the sample of low-mass KOIs.  We use the $K$-band indices to measure $T_{\rm eff}$ and [M/H], then interpolate those values on the 5-Gyr Dartmouth isochrones to determine stellar mass and stellar radius \citep{Dotter2008, Feiden2011}.  {\it Top}: Stellar radius vs. $T_{\rm eff}$.  Squares with errorbars are the measurements in this paper and black diamonds are the values from the KIC.  {\it Bot}: The difference between the radius determinations from the KIC and in this work, plotted against our $T_{\rm eff}$ measurements.  We dramatically revise the stellar radii of the coolest KOIs.  KOI 961 is cooler than the available grid points in the Dartmouth isochrones, so we instead interpolate the $T_{\rm eff}$ and [M/H] onto the 5 Gyr BCAH isochrones \citet{Baraffe1998}.  KOI 961 is the subject of the third paper in this paper series \citep{Muirhead2012}, wherein we use Barnard's Star to more accurately estimate its stellar parameters and refit the transit parameters.}
\label{fig:radii} 
\end{center}
\end{figure}

All but one of the stars in our sample are treated homogeneously, so the relative radii, masses and temperatures should be precise, even in the presence of model-dependent offsets.  There is no model in the Dartmouth isochrones with the same metallicity and effective temperature as KOI 961, so this star is interpolated onto the 5-Gyr BCAH isochrones \citep[see ][ however, for a more detailed analysis of this star]{Muirhead2012}.  Stellar masses and radii are calculated by interpolation of the main sequence of a 5-Gyr isochrone at the measured total metallicity and effective temperature, illustrated in Figure \ref{fig:radii}.  The assumption of age does not significantly change the results.  If very young ages are adopted, the masses and radii typically change only 0.1\% between ages of 1 and 10 Gyr, and in all cases significantly less than the reported uncertainties.

We also apply our method to stars with $K$-band measurements in \citet{Rojas2012} and radius measurements using optical-long baseline interferometry (see Figure \ref{comparison}, Panels B and C).  We find good agreement to within the estimated uncertainties.

\section{Error Analysis}\label{errors}

We estimate the uncertainty in the equivalent width, $\rm H_2O$-K2 index and [M/H] measurements due to noise in the spectra using Monte Carlo simulations.  Spextool reports errors for each channel of a reduced spectrum assuming photon noise and read noise in the target and telluric calibrator exposures.  For each reduced spectrum, we created 1000 copies, each with random noise added to the spectral channels based on the error reported by Spextool.  For each of the 1000 simulations, we measure the Na I and Ca I equivalent widths and $\rm H_2O$-K2 index.  We also calculate [M/H] and $T_{\rm eff}$, and interpolate those values onto the Dartmouth isochrones.  The standard deviations of the quantities across the simulations are taken as the uncertainty in those quantities for a given KOI.  All of the resulting distributions are reasonably symmetric, except for the $T_{\rm eff}$ measurements, for which we report asymetric uncertainties.

The [M/H] and $T_{\rm eff}$ measurements contain additional uncertainty from imperfections in the calibration relation.  Possible sources of calibration errors include astrophysical scatter from non-perfect correlation between the indices and [M/H], as well as errors in the relation coefficients due to noise in the calibration spectra.  \citet{Rojas2012} estimated the calibration errors contribute 0.1 dex of error to [M/H] measurements, based on the root-mean-square residuals in the calibration fit.  We estimate a $T_{\rm eff}$ calibration uncertainty of 50 K by comparing the method to OLBI measurements (see Figure \ref{comparison}).  We include these uncertainties into the $M_\star$ and $R_\star$ uncertainty estimates by adding additional Gaussian noise into the [M/H] and $T_{\rm eff}$ measurements of each Monte Carlo iteration with standard deviations of 0.1 and 50 K respectively.

A final uncertainty term arrises from the inherent accuracy of the Dartmouth evolutionary isochrones.  Figure \ref{comparison} shows $R_\star$ as measured by optical long-baseline interferometry (OLBI) and $M_\star$ as measured using the \citet{Delfosse2000} mass-$M_K$ relation, versus measurements made using the Dartmouth interpolations for nearby stars with overlapping measurements in \citep{Rojas2012}.  We see no evidence for systematic offsets in the Dartmouth interpolations; nevertheless, we add in quadrature an additional uncertainty corresponding to 10\% of the interpolated $M_\star$ and $R_\star$ to the Monte Carlo uncertainties based on the residual differences.

\section{Results and Discussion}\label{discussion}

Table 1 lists the KOI planet candidates and our measurements of their host star paramaters: spectral type, effective temperature, metallicity, and mass and radius, with corresponding uncertainty estimates.  We also include our estimate of the the planet candidate radii, calculated by applying the $R_P / R_\star$ measurements reported in \citet{Borucki2011} to our measurements of $R_\star$.   It should be noted that the appropriate limb-darkening coefficient could change as a result of the stellar classification, requiring a more sophisticated calculation of the new planet-candidate radii than in this letter.

Figure \ref{fig:radii} plots the effective temperatures, metallicities, masses and radii from our analysis as well as the values in the KIC.  Our stellar radii are systematically lower than the values reported in the KIC and, by construction, have better agreement with $T_{\rm eff}$.  The smaller stellar radii imply smaller planet-candidate radii, and many of the revised planet-candidate radii are smaller than 1 Earth radius.

The effective temperatures, radii and masses of the KOIs imply different planet-candidate equilibrium temperature estimates, such that 3 planet-candidates are terrestrial-sized and have equilibrium temperatures which may permit liquid water to reside on the planet surface, assuming Earth-like albedos and re-radiation fractions.  We find that KOIs 463.01, 812.03, and 854.01 are less than 2 $R_\Earth$ in size with equilibrium temperatures between 215 K and 275 K: rough limits to the habitable zone as calculated by \citet{Kasting1993}.  Whether liquid water can persist on the surfaces of terrestrial planets orbiting low-mass stars depends strongly on the individual evolution and atmospheric peculiarities of each system, but these KOIs are nevertheless compelling targets for future followup work.

A recent paper by \citet{Gaidos2012} compares the statistics from the M2K M-dwarf Doppler survey \citep[e.g.][]{Apps2010} with the Kepler results.  They found inconsistencies which could be explained if many of the KOI stellar radii were underestimated, a result which is contradictory to our findings.  However, the KIC $T_{\rm eff}$ used to compare the M2K and Kepler samples differ from our measurements, and the two samples may have different metallicity distributions.

\acknowledgements

We would like to thank Michael Cushing for providing us with a version of the Spextool package for the TripleSpec Spectrograph at Palomar.  We would also like to thank John Johnson for his thoughtful comments on the letter.  The Palomar 200-inch Telescope time was provided by Cornell University.  K.R.C. acknowledges support for this work from the Hubble Fellowship Program, provided by NASA through Hubble Fellowship grant HST-HF-51253.01-A awarded by the STScI, which is operated by the AURA, Inc., for NASA, under contract NAS 5-26555.


\LongTables
\clearpage
\begin{landscape}
\begin{deluxetable}{rcccccccccccc}
\tablewidth{0pt}
\tabletypesize{\scriptsize}
\tablecaption{Cool KOI Stellar Host and Planet Candidate Properties  \label{tab:KOIproperties}}
\tablehead{
\multicolumn{1}{c}{} & \multicolumn{2}{c}{-- \citet{Borucki2011} --} & \multicolumn{3}{c}{-- $K$-Band Stellar Measurements --} & \multicolumn{2}{c}{-- Dartmouth Stellar Interpolants --} &  \multicolumn{3}{c}{-- New Planet Parameters --} \\\multicolumn{1}{c}{} & \multicolumn{2}{c}{Transit Parameters} & \multicolumn{3}{c}{(This Work)} & \multicolumn{2}{c}{(This Work)} & \multicolumn{3}{c}{(This Work)} \\ \\
  \colhead{KOI} &
  \colhead{$P$ [days]} &
  \colhead{$R_{\rm P}/R_{\star}$} &
  \colhead{$T_{\rm eff}$ [K]} &
  \colhead{[M/H]\tablenotemark{a}} &
  \colhead{Sp. Type} &
  \colhead{$M_{\star}$ [$M_\Sun$]} &
  \colhead{$R_{\star}$ [$R_\Sun$]} &
  \colhead{$a$ [AU]} &
  \colhead{$R_{\rm P}$ [$R_\Earth$]} &
\colhead{$T_{\rm eq,P}$ [K]\tablenotemark{b}}}
\startdata
104.01 & 2.5080910 $\pm$ 1.3e-05 & 0.035 $\pm$ 4.8e-03 & $4238^{+459}_{-106}$ & -0.12 $\pm$ 0.10 & M0V & 0.64 $\pm$ 0.08 & 0.61 $\pm$ 0.08 & 0.03106 $\pm$ 1.29e-03 & 2.32 $\pm$ 0.43 & 827\\
156.01 & 8.0414400 $\pm$ 1.3e-04 & 0.023 $\pm$ 1.3e-02 & $3983^{+80}_{-81}$ & -0.20 $\pm$ 0.10 & M0V & 0.56 $\pm$ 0.06 & 0.54 $\pm$ 0.06 & 0.06482 $\pm$ 2.35e-03 & 1.36 $\pm$ 0.78 & 508\\
156.02 & 5.1885600 $\pm$ 1.2e-04 & 0.020 $\pm$ 1.4e-02 & ... & ... & ... & ... & ... & 0.04840 $\pm$ 1.75e-03 & 1.18 $\pm$ 0.84 & 588\\
156.03 & 11.7761793 $\pm$ 5.4e-05 & 0.033 $\pm$ 2.8e-04 & ... & ... & ... & ... & ... & 0.08358 $\pm$ 3.03e-03 & 1.95 $\pm$ 0.22 & 447\\
222.01 & 6.3123822 $\pm$ 5.8e-05 & 0.033 $\pm$ 3.9e-04 & $4096^{+538}_{-187}$ & -0.17 $\pm$ 0.11 & M0V & 0.59 $\pm$ 0.09 & 0.57 $\pm$ 0.09 & 0.05611 $\pm$ 2.81e-03 & 2.05 $\pm$ 0.31 & 576\\
222.02 & 12.7939701 $\pm$ 3.0e-04 & 0.026 $\pm$ 6.5e-04 & ... & ... & ... & ... & ... & 0.08987 $\pm$ 4.50e-03 & 1.62 $\pm$ 0.25 & 455\\
227.01 & 17.6607609 $\pm$ 2.5e-04 & 0.040 $\pm$ 3.9e-04 & $3745^{+51}_{-59}$ & -0.02 $\pm$ 0.10 & M1V & 0.49 $\pm$ 0.06 & 0.47 $\pm$ 0.05 & 0.10496 $\pm$ 3.91e-03 & 2.06 $\pm$ 0.24 & 350\\
247.01 & 13.8152399 $\pm$ 3.2e-04 & 0.031 $\pm$ 1.1e-03 & $3741^{+54}_{-51}$ & 0.03 $\pm$ 0.10 & M1V & 0.51 $\pm$ 0.06 & 0.49 $\pm$ 0.06 & 0.08992 $\pm$ 3.29e-03 & 1.64 $\pm$ 0.20 & 383\\
248.01 & 7.2034941 $\pm$ 6.5e-05 & 0.039 $\pm$ 5.5e-04 & $3816^{+141}_{-54}$ & -0.05 $\pm$ 0.11 & M1V & 0.54 $\pm$ 0.06 & 0.51 $\pm$ 0.06 & 0.05929 $\pm$ 2.27e-03 & 2.18 $\pm$ 0.27 & 495\\
248.02 & 10.9140100 $\pm$ 1.8e-04 & 0.034 $\pm$ 1.1e-03 & ... & ... & ... & ... & ... & 0.07821 $\pm$ 3.00e-03 & 1.90 $\pm$ 0.24 & 431\\
248.03 & 2.5765359 $\pm$ 3.3e-05 & 0.027 $\pm$ 1.2e-02 & ... & ... & ... & ... & ... & 0.02987 $\pm$ 1.14e-03 & 1.51 $\pm$ 0.70 & 697\\
250.01 & 12.2823563 $\pm$ 9.7e-05 & 0.049 $\pm$ 4.8e-04 & $3887^{+258}_{-57}$ & -0.08 $\pm$ 0.11 & M0V & 0.55 $\pm$ 0.08 & 0.53 $\pm$ 0.08 & 0.08551 $\pm$ 3.93e-03 & 2.83 $\pm$ 0.40 & 427\\
250.02 & 17.2520409 $\pm$ 2.5e-04 & 0.049 $\pm$ 1.5e-03 & ... & ... & ... & ... & ... & 0.10725 $\pm$ 4.93e-03 & 2.83 $\pm$ 0.41 & 381\\
250.03 & 3.5438709 $\pm$ 7.1e-05 & 0.017 $\pm$ 1.0e-03 & ... & ... & ... & ... & ... & 0.03734 $\pm$ 1.72e-03 & 0.98 $\pm$ 0.15 & 646\\
251.01 & 4.1643710 $\pm$ 2.3e-05 & 0.044 $\pm$ 5.0e-04 & $3810^{+64}_{-65}$ & -0.02 $\pm$ 0.10 & M1V & 0.54 $\pm$ 0.06 & 0.52 $\pm$ 0.06 & 0.04124 $\pm$ 1.51e-03 & 2.47 $\pm$ 0.28 & 594\\
252.01 & 17.6043892 $\pm$ 2.4e-04 & 0.042 $\pm$ 2.2e-02 & $3743^{+63}_{-65}$ & 0.06 $\pm$ 0.11 & M1V & 0.51 $\pm$ 0.06 & 0.49 $\pm$ 0.06 & 0.10593 $\pm$ 4.00e-03 & 2.24 $\pm$ 1.20 & 354\\
253.01 & 6.3832402 $\pm$ 7.1e-05 & 0.043 $\pm$ 7.9e-04 & $3769^{+297}_{-50}$ & 0.40 $\pm$ 0.13 & M1V & 0.59 $\pm$ 0.08 & 0.56 $\pm$ 0.08 & 0.05633 $\pm$ 2.49e-03 & 2.62 $\pm$ 0.36 & 524\\
254.01 & 2.4552391 $\pm$ 1.6e-06 & 0.184 $\pm$ 1.2e-03 & $3814^{+124}_{-54}$ & 0.20 $\pm$ 0.11 & M1V & 0.58 $\pm$ 0.07 & 0.55 $\pm$ 0.06 & 0.02967 $\pm$ 1.12e-03 & 11.11 $\pm$ 1.30 & 727\\
255.01 & 27.5215607 $\pm$ 3.1e-04 & 0.045 $\pm$ 4.7e-04 & $3770^{+66}_{-78}$ & 0.02 $\pm$ 0.11 & M1V & 0.53 $\pm$ 0.06 & 0.51 $\pm$ 0.06 & 0.14469 $\pm$ 5.52e-03 & 2.50 $\pm$ 0.30 & 312\\
256.01 & 1.3786809 $\pm$ 1.2e-05 & 0.123 $\pm$ 2.3e-03 & $3450^{+50}_{-58}$ & 0.31 $\pm$ 0.10 & M3V & 0.43 $\pm$ 0.06 & 0.42 $\pm$ 0.06 & 0.01833 $\pm$ 8.10e-04 & 5.60 $\pm$ 0.76 & 726\\
314.01 & 13.7810497 $\pm$ 1.6e-04 & 0.029 $\pm$ 5.8e-04 & $3841^{+50}_{-51}$ & -0.18 $\pm$ 0.10 & M1V & 0.52 $\pm$ 0.06 & 0.50 $\pm$ 0.06 & 0.09060 $\pm$ 3.24e-03 & 1.58 $\pm$ 0.18 & 398\\
314.02 & 23.0904007 $\pm$ 3.4e-04 & 0.023 $\pm$ 3.9e-02 & ... & ... & ... & ... & ... & 0.12781 $\pm$ 4.56e-03 & 1.26 $\pm$ 2.13 & 335\\
430.01 & 12.3764496 $\pm$ 1.9e-04 & 0.038 $\pm$ 1.8e-02 & $3797^{+98}_{-58}$ & -0.10 $\pm$ 0.11 & M1V & 0.51 $\pm$ 0.06 & 0.48 $\pm$ 0.06 & 0.08346 $\pm$ 3.25e-03 & 2.00 $\pm$ 0.98 & 403\\
438.01 & 5.9312038 $\pm$ 7.0e-05 & 0.030 $\pm$ 1.6e-02 & $3985^{+251}_{-64}$ & -0.22 $\pm$ 0.11 & M0V & 0.56 $\pm$ 0.08 & 0.54 $\pm$ 0.07 & 0.05288 $\pm$ 2.27e-03 & 1.76 $\pm$ 0.97 & 560\\
448.01 & 10.1396103 $\pm$ 2.3e-04 & 0.030 $\pm$ 9.4e-04 & $3894^{+67}_{-98}$ & -0.28 $\pm$ 0.11 & M0V & 0.52 $\pm$ 0.06 & 0.50 $\pm$ 0.06 & 0.07369 $\pm$ 2.84e-03 & 1.64 $\pm$ 0.21 & 448\\
448.02 & 43.6204987 $\pm$ 1.3e-03 & 0.049 $\pm$ 8.8e-03 & ... & ... & ... & ... & ... & 0.19491 $\pm$ 7.52e-03 & 2.68 $\pm$ 0.58 & 275\\
463.01 & 18.4781704 $\pm$ 1.9e-04 & 0.047 $\pm$ 9.2e-04 & $3387^{+59}_{-50}$ & -0.09 $\pm$ 0.10 & M3V & 0.30 $\pm$ 0.05 & 0.30 $\pm$ 0.05 & 0.09116 $\pm$ 5.22e-03 & 1.52 $\pm$ 0.26 & 269\\
478.01 & 11.0234518 $\pm$ 8.5e-05 & 0.051 $\pm$ 1.8e-03 & $3744^{+50}_{-61}$ & 0.13 $\pm$ 0.10 & M1V & 0.52 $\pm$ 0.06 & 0.50 $\pm$ 0.06 & 0.07812 $\pm$ 2.80e-03 & 2.78 $\pm$ 0.33 & 418\\
494.01 & 25.6971893 $\pm$ 6.7e-04 & 0.031 $\pm$ 2.4e-02 & $3789^{+219}_{-159}$ & -0.21 $\pm$ 0.16 & M1V & 0.50 $\pm$ 0.10 & 0.48 $\pm$ 0.10 & 0.13495 $\pm$ 8.57e-03 & 1.61 $\pm$ 1.29 & 314\\
500.01 & 7.0534778 $\pm$ 8.3e-05 & 0.034 $\pm$ 5.4e-04 & $4040^{+64}_{-175}$ & -0.14 $\pm$ 0.11 & M0V & 0.59 $\pm$ 0.07 & 0.57 $\pm$ 0.07 & 0.06022 $\pm$ 2.34e-03 & 2.10 $\pm$ 0.26 & 546\\
500.02 & 9.5216999 $\pm$ 1.4e-04 & 0.035 $\pm$ 2.3e-02 & ... & ... & ... & ... & ... & 0.07355 $\pm$ 2.86e-03 & 2.16 $\pm$ 1.44 & 494\\
500.03 & 3.0721660 $\pm$ 4.6e-05 & 0.018 $\pm$ 7.4e-04 & ... & ... & ... & ... & ... & 0.03460 $\pm$ 1.34e-03 & 1.11 $\pm$ 0.14 & 720\\
500.04 & 4.6453528 $\pm$ 6.7e-05 & 0.026 $\pm$ 2.3e-02 & ... & ... & ... & ... & ... & 0.04558 $\pm$ 1.77e-03 & 1.60 $\pm$ 1.43 & 628\\
500.05 & 0.9867790 $\pm$ 1.2e-05 & 0.015 $\pm$ 1.4e-02 & ... & ... & ... & ... & ... & 0.01623 $\pm$ 6.30e-04 & 0.93 $\pm$ 0.87 & 1052\\
503.01 & 8.2223597 $\pm$ 1.0e-04 & 0.034 $\pm$ 6.5e-04 & $4003^{+179}_{-75}$ & -0.13 $\pm$ 0.11 & M0V & 0.58 $\pm$ 0.07 & 0.56 $\pm$ 0.07 & 0.06633 $\pm$ 2.63e-03 & 2.06 $\pm$ 0.25 & 511\\
531.01 & 3.6874621 $\pm$ 8.9e-06 & 0.055 $\pm$ 7.6e-03 & $4030^{+82}_{-169}$ & 0.10 $\pm$ 0.10 & M0V & 0.62 $\pm$ 0.08 & 0.60 $\pm$ 0.07 & 0.03990 $\pm$ 1.55e-03 & 3.59 $\pm$ 0.66 & 689\\
571.01 & 7.2673302 $\pm$ 1.3e-04 & 0.024 $\pm$ 1.5e-02 & $3761^{+77}_{-50}$ & -0.21 $\pm$ 0.11 & M1V & 0.48 $\pm$ 0.06 & 0.46 $\pm$ 0.06 & 0.05763 $\pm$ 2.23e-03 & 1.21 $\pm$ 0.77 & 470\\
571.02 & 13.3433104 $\pm$ 1.5e-04 & 0.028 $\pm$ 6.7e-04 & ... & ... & ... & ... & ... & 0.08642 $\pm$ 3.34e-03 & 1.41 $\pm$ 0.17 & 384\\
571.03 & 3.8867581 $\pm$ 3.8e-05 & 0.026 $\pm$ 7.0e-03 & ... & ... & ... & ... & ... & 0.03797 $\pm$ 1.47e-03 & 1.31 $\pm$ 0.39 & 579\\
596.01 & 1.6827060 $\pm$ 1.4e-05 & 0.026 $\pm$ 4.2e-04 & $3678^{+52}_{-60}$ & 0.00 $\pm$ 0.10 & M1V & 0.49 $\pm$ 0.06 & 0.47 $\pm$ 0.06 & 0.02188 $\pm$ 8.60e-04 & 1.34 $\pm$ 0.16 & 753\\
605.01 & 2.6281440 $\pm$ 2.0e-05 & 0.028 $\pm$ 4.4e-04 & $3941^{+369}_{-73}$ & -0.03 $\pm$ 0.16 & M0V & 0.58 $\pm$ 0.09 & 0.56 $\pm$ 0.09 & 0.03101 $\pm$ 1.56e-03 & 1.70 $\pm$ 0.26 & 736\\
610.01 & 14.2824602 $\pm$ 2.6e-04 & 0.027 $\pm$ 8.4e-04 & $3679^{+50}_{-64}$ & -0.02 $\pm$ 0.10 & M1V & 0.49 $\pm$ 0.06 & 0.47 $\pm$ 0.06 & 0.09066 $\pm$ 3.60e-03 & 1.37 $\pm$ 0.17 & 368\\
663.01 & 2.7556019 $\pm$ 1.7e-05 & 0.025 $\pm$ 6.5e-03 & $3834^{+50}_{-57}$ & -0.25 $\pm$ 0.10 & M1V & 0.51 $\pm$ 0.06 & 0.48 $\pm$ 0.06 & 0.03068 $\pm$ 1.12e-03 & 1.32 $\pm$ 0.38 & 672\\
663.02 & 20.3070793 $\pm$ 2.9e-04 & 0.023 $\pm$ 4.3e-04 & ... & ... & ... & ... & ... & 0.11617 $\pm$ 4.26e-03 & 1.22 $\pm$ 0.14 & 345\\
676.01 & 7.9725132 $\pm$ 7.6e-05 & 0.059 $\pm$ 3.4e-03 & $3914^{+82}_{-59}$ & -0.12 $\pm$ 0.10 & M0V & 0.55 $\pm$ 0.06 & 0.53 $\pm$ 0.06 & 0.06406 $\pm$ 2.29e-03 & 3.39 $\pm$ 0.43 & 495\\
676.02 & 2.4532239 $\pm$ 2.3e-05 & 0.039 $\pm$ 8.6e-04 & ... & ... & ... & ... & ... & 0.02920 $\pm$ 1.04e-03 & 2.24 $\pm$ 0.26 & 734\\
736.01 & 18.7952309 $\pm$ 6.0e-04 & 0.036 $\pm$ 3.8e-02 & $3682^{+103}_{-102}$ & -0.11 $\pm$ 0.14 & M1V & 0.46 $\pm$ 0.08 & 0.44 $\pm$ 0.07 & 0.10714 $\pm$ 5.54e-03 & 1.74 $\pm$ 1.86 & 330\\
736.02 & 6.7388001 $\pm$ 2.1e-04 & 0.026 $\pm$ 1.1e-03 & ... & ... & ... & ... & ... & 0.05407 $\pm$ 2.80e-03 & 1.26 $\pm$ 0.21 & 465\\
739.01 & 1.2870520 $\pm$ 1.2e-05 & 0.027 $\pm$ 5.1e-04 & $3733^{+76}_{-59}$ & 0.07 $\pm$ 0.11 & M1V & 0.53 $\pm$ 0.06 & 0.51 $\pm$ 0.06 & 0.01873 $\pm$ 7.00e-04 & 1.49 $\pm$ 0.18 & 856\\
775.01 & 16.3852291 $\pm$ 7.0e-04 & 0.029 $\pm$ 1.6e-03 & $3898^{+177}_{-77}$ & -0.09 $\pm$ 0.12 & M0V & 0.56 $\pm$ 0.07 & 0.54 $\pm$ 0.07 & 0.10379 $\pm$ 4.44e-03 & 1.69 $\pm$ 0.24 & 390\\
775.02 & 7.8776102 $\pm$ 1.8e-04 & 0.033 $\pm$ 1.2e-03 & ... & ... & ... & ... & ... & 0.06370 $\pm$ 2.72e-03 & 1.93 $\pm$ 0.27 & 499\\
778.01 & 2.2433400 $\pm$ 3.6e-05 & 0.028 $\pm$ 1.0e-03 & $3741^{+140}_{-54}$ & -0.16 $\pm$ 0.12 & M1V & 0.47 $\pm$ 0.06 & 0.45 $\pm$ 0.06 & 0.02612 $\pm$ 1.10e-03 & 1.38 $\pm$ 0.19 & 686\\
781.01 & 11.5982304 $\pm$ 1.5e-04 & 0.050 $\pm$ 2.9e-02 & $3672^{+103}_{-58}$ & 0.09 $\pm$ 0.12 & M1V & 0.51 $\pm$ 0.06 & 0.49 $\pm$ 0.06 & 0.08015 $\pm$ 3.24e-03 & 2.66 $\pm$ 1.58 & 399\\
784.01 & 19.2693005 $\pm$ 5.9e-04 & 0.032 $\pm$ 3.3e-02 & $3767^{+135}_{-51}$ & -0.10 $\pm$ 0.11 & M1V & 0.51 $\pm$ 0.06 & 0.48 $\pm$ 0.06 & 0.11210 $\pm$ 4.43e-03 & 1.69 $\pm$ 1.76 & 345\\
812.01 & 3.3402400 $\pm$ 3.5e-05 & 0.039 $\pm$ 2.8e-02 & $3887^{+188}_{-78}$ & -0.36 $\pm$ 0.12 & M0V & 0.51 $\pm$ 0.07 & 0.48 $\pm$ 0.07 & 0.03487 $\pm$ 1.57e-03 & 2.06 $\pm$ 1.50 & 638\\
812.02 & 20.0608597 $\pm$ 4.7e-04 & 0.035 $\pm$ 1.1e-03 & ... & ... & ... & ... & ... & 0.11521 $\pm$ 5.18e-03 & 1.84 $\pm$ 0.27 & 351\\
812.03 & 46.1851006 $\pm$ 1.9e-03 & 0.034 $\pm$ 1.7e-03 & ... & ... & ... & ... & ... & 0.20087 $\pm$ 9.03e-03 & 1.79 $\pm$ 0.27 & 266\\
817.01 & 23.9715996 $\pm$ 1.1e-03 & 0.033 $\pm$ 2.0e-03 & $3750^{+66}_{-60}$ & 0.08 $\pm$ 0.11 & M1V & 0.54 $\pm$ 0.06 & 0.51 $\pm$ 0.06 & 0.13215 $\pm$ 4.89e-03 & 1.84 $\pm$ 0.24 & 325\\
818.01 & 8.1142902 $\pm$ 1.3e-04 & 0.040 $\pm$ 2.1e-02 & $3721^{+50}_{-118}$ & 0.19 $\pm$ 0.12 & M1V & 0.54 $\pm$ 0.07 & 0.52 $\pm$ 0.06 & 0.06455 $\pm$ 2.56e-03 & 2.27 $\pm$ 1.22 & 466\\
854.01 & 56.0517006 $\pm$ 2.7e-03 & 0.036 $\pm$ 1.6e-03 & $3593^{+58}_{-66}$ & 0.16 $\pm$ 0.11 & M2V & 0.49 $\pm$ 0.06 & 0.47 $\pm$ 0.06 & 0.22599 $\pm$ 9.31e-03 & 1.83 $\pm$ 0.25 & 227\\
868.01 & 234.0000000 $\pm$ 1.4e+01 & 0.161 $\pm$ 2.5e-03 & $3822^{+177}_{-58}$ & 0.02 $\pm$ 0.12 & M1V & 0.55 $\pm$ 0.07 & 0.53 $\pm$ 0.07 & 0.60896 $\pm$ 2.50e-02 & 9.27 $\pm$ 1.20 & 157\\
870.01 & 5.9121299 $\pm$ 1.7e-04 & 0.030 $\pm$ 1.3e-03 & $4072^{+720}_{-133}$ & -0.17 $\pm$ 0.11 & M0V & 0.59 $\pm$ 0.10 & 0.57 $\pm$ 0.09 & 0.05375 $\pm$ 2.91e-03 & 1.86 $\pm$ 0.32 & 583\\
870.02 & 8.9859695 $\pm$ 2.7e-04 & 0.028 $\pm$ 8.9e-04 & ... & ... & ... & ... & ... & 0.07106 $\pm$ 3.84e-03 & 1.73 $\pm$ 0.29 & 507\\
875.01 & 4.2209358 $\pm$ 2.7e-05 & 0.044 $\pm$ 3.4e-04 & $3861^{+739}_{-57}$ & -0.05 $\pm$ 0.13 & M0V & 0.55 $\pm$ 0.10 & 0.52 $\pm$ 0.09 & 0.04183 $\pm$ 2.32e-03 & 2.51 $\pm$ 0.43 & 602\\
877.01 & 5.9548702 $\pm$ 1.0e-04 & 0.034 $\pm$ 2.6e-02 & $3896^{+237}_{-50}$ & -0.20 $\pm$ 0.11 & M0V & 0.54 $\pm$ 0.07 & 0.52 $\pm$ 0.07 & 0.05220 $\pm$ 2.21e-03 & 1.92 $\pm$ 1.49 & 541\\
877.02 & 12.0395698 $\pm$ 3.4e-04 & 0.031 $\pm$ 1.1e-03 & ... & ... & ... & ... & ... & 0.08347 $\pm$ 3.53e-03 & 1.75 $\pm$ 0.24 & 428\\
886.01 & 8.0128098 $\pm$ 1.6e-04 & 0.035 $\pm$ 4.0e-02 & $3726^{+69}_{-71}$ & -0.12 $\pm$ 0.11 & M1V & 0.49 $\pm$ 0.06 & 0.47 $\pm$ 0.06 & 0.06190 $\pm$ 2.46e-03 & 1.80 $\pm$ 2.07 & 453\\
898.01 & 9.7705898 $\pm$ 1.8e-04 & 0.041 $\pm$ 3.7e-02 & $3893^{+102}_{-95}$ & -0.15 $\pm$ 0.11 & M0V & 0.54 $\pm$ 0.07 & 0.52 $\pm$ 0.06 & 0.07295 $\pm$ 2.84e-03 & 2.34 $\pm$ 2.13 & 460\\
898.02 & 5.1699100 $\pm$ 1.1e-04 & 0.033 $\pm$ 3.2e-02 & ... & ... & ... & ... & ... & 0.04773 $\pm$ 1.86e-03 & 1.88 $\pm$ 1.84 & 569\\
898.03 & 20.0892296 $\pm$ 5.4e-04 & 0.034 $\pm$ 4.1e-02 & ... & ... & ... & ... & ... & 0.11796 $\pm$ 4.59e-03 & 1.94 $\pm$ 2.35 & 362\\
899.01 & 7.1138802 $\pm$ 1.1e-04 & 0.028 $\pm$ 1.9e-02 & $3568^{+64}_{-51}$ & 0.02 $\pm$ 0.10 & M2V & 0.44 $\pm$ 0.06 & 0.42 $\pm$ 0.06 & 0.05504 $\pm$ 2.38e-03 & 1.28 $\pm$ 0.89 & 435\\
899.02 & 3.3065691 $\pm$ 3.9e-05 & 0.021 $\pm$ 6.6e-04 & ... & ... & ... & ... & ... & 0.03303 $\pm$ 1.43e-03 & 0.96 $\pm$ 0.13 & 562\\
899.03 & 15.3681297 $\pm$ 2.6e-04 & 0.029 $\pm$ 3.6e-02 & ... & ... & ... & ... & ... & 0.09198 $\pm$ 3.98e-03 & 1.33 $\pm$ 1.66 & 336\\
901.01 & 12.7325573 $\pm$ 4.6e-05 & 0.074 $\pm$ 5.4e-04 & $3945^{+811}_{-50}$ & -0.14 $\pm$ 0.11 & M0V & 0.56 $\pm$ 0.10 & 0.54 $\pm$ 0.09 & 0.08792 $\pm$ 4.87e-03 & 4.36 $\pm$ 0.73 & 431\\
902.01 & 83.9041977 $\pm$ 1.7e-03 & 0.080 $\pm$ 6.6e-04 & $3960^{+140}_{-133}$ & -0.37 $\pm$ 0.12 & M0V & 0.53 $\pm$ 0.07 & 0.51 $\pm$ 0.07 & 0.30300 $\pm$ 1.34e-02 & 4.41 $\pm$ 0.62 & 225\\
912.01 & 10.8484697 $\pm$ 2.7e-04 & 0.037 $\pm$ 3.1e-02 & $4208^{+1067}_{-111}$ & -0.19 $\pm$ 0.11 & M0V & 0.61 $\pm$ 0.10 & 0.59 $\pm$ 0.09 & 0.08145 $\pm$ 4.33e-03 & 2.40 $\pm$ 2.04 & 501\\
936.01 & 9.4678946 $\pm$ 8.3e-05 & 0.044 $\pm$ 4.6e-04 & $3581^{+65}_{-50}$ & 0.14 $\pm$ 0.10 & M2V & 0.46 $\pm$ 0.06 & 0.44 $\pm$ 0.06 & 0.06753 $\pm$ 2.77e-03 & 2.12 $\pm$ 0.27 & 404\\
936.02 & 0.8930440 $\pm$ 4.4e-06 & 0.026 $\pm$ 1.1e-02 & ... & ... & ... & ... & ... & 0.01399 $\pm$ 5.70e-04 & 1.25 $\pm$ 0.55 & 887\\
947.01 & 28.5989094 $\pm$ 5.7e-04 & 0.040 $\pm$ 9.3e-04 & $3750^{+56}_{-75}$ & -0.18 $\pm$ 0.11 & M1V & 0.49 $\pm$ 0.06 & 0.46 $\pm$ 0.06 & 0.14388 $\pm$ 5.66e-03 & 2.02 $\pm$ 0.25 & 296\\
952.01 & 5.9012551 $\pm$ 7.6e-05 & 0.037 $\pm$ 6.7e-04 & $3727^{+104}_{-64}$ & 0.04 $\pm$ 0.13 & M1V & 0.52 $\pm$ 0.07 & 0.50 $\pm$ 0.06 & 0.05151 $\pm$ 2.06e-03 & 2.02 $\pm$ 0.26 & 512\\
952.02 & 8.7524595 $\pm$ 1.5e-04 & 0.038 $\pm$ 1.1e-02 & ... & ... & ... & ... & ... & 0.06699 $\pm$ 2.68e-03 & 2.08 $\pm$ 0.65 & 449\\
952.03 & 22.7803307 $\pm$ 3.1e-04 & 0.040 $\pm$ 9.3e-04 & ... & ... & ... & ... & ... & 0.12676 $\pm$ 5.07e-03 & 2.18 $\pm$ 0.28 & 326\\
952.04 & 2.8960290 $\pm$ 6.3e-05 & 0.019 $\pm$ 1.5e-03 & ... & ... & ... & ... & ... & 0.03205 $\pm$ 1.28e-03 & 1.04 $\pm$ 0.15 & 650\\
961.01\tablenotemark{c} & 1.2137721 $\pm$ 4.4e-06 & 0.053 $\pm$ 3.3e-03 & $3203^{+56}_{-51}$ & -0.33 $\pm$ 0.10 & M4V & 0.14 $\pm$ 0.02 & 0.16 $\pm$ 0.02 & 0.01153 $\pm$ 5.50e-04 & 0.92 $\pm$ 0.14 & 526\\
961.02 & 0.4532880 $\pm$ 9.0e-07 & 0.194 $\pm$ 8.1e-04 & ... & ... & ... & ... & ... & 0.00598 $\pm$ 2.80e-04 & 3.38 $\pm$ 0.46 & 730\\
961.03 & 1.8651130 $\pm$ 9.1e-06 & 0.140 $\pm$ 2.3e-01 & ... & ... & ... & ... & ... & 0.01535 $\pm$ 7.30e-04 & 2.44 $\pm$ 4.02 & 456\\
1024.01 & 5.7477322 $\pm$ 6.5e-05 & 0.026 $\pm$ 8.2e-04 & $3937^{+163}_{-61}$ & -0.27 $\pm$ 0.11 & M0V & 0.53 $\pm$ 0.07 & 0.52 $\pm$ 0.06 & 0.05097 $\pm$ 1.99e-03 & 1.47 $\pm$ 0.19 & 554\\
1026.01 & 94.1023026 $\pm$ 9.7e-03 & 0.024 $\pm$ 8.0e-04 & $3773^{+58}_{-61}$ & 0.03 $\pm$ 0.10 & M1V & 0.54 $\pm$ 0.06 & 0.51 $\pm$ 0.06 & 0.32898 $\pm$ 1.20e-02 & 1.34 $\pm$ 0.16 & 208\\
1078.01 & 3.3536820 $\pm$ 2.9e-05 & 0.035 $\pm$ 1.0e-03 & $3807^{+97}_{-69}$ & -0.26 $\pm$ 0.15 & M1V & 0.49 $\pm$ 0.07 & 0.47 $\pm$ 0.06 & 0.03460 $\pm$ 1.49e-03 & 1.79 $\pm$ 0.25 & 617\\
1085.01 & 7.7179398 $\pm$ 2.3e-04 & 0.017 $\pm$ 1.9e-03 & $3939^{+231}_{-86}$ & -0.25 $\pm$ 0.12 & M0V & 0.54 $\pm$ 0.08 & 0.52 $\pm$ 0.07 & 0.06219 $\pm$ 2.78e-03 & 0.97 $\pm$ 0.17 & 503\\
1141.01 & 5.7279601 $\pm$ 1.5e-04 & 0.027 $\pm$ 1.2e-03 & $4065^{+577}_{-137}$ & -0.13 $\pm$ 0.11 & M0V & 0.59 $\pm$ 0.09 & 0.57 $\pm$ 0.09 & 0.05268 $\pm$ 2.65e-03 & 1.69 $\pm$ 0.27 & 591\\
1146.01 & 7.0969501 $\pm$ 2.0e-04 & 0.024 $\pm$ 4.4e-02 & $3555^{+83}_{-52}$ & -0.07 $\pm$ 0.11 & M2V & 0.41 $\pm$ 0.06 & 0.40 $\pm$ 0.06 & 0.05383 $\pm$ 2.51e-03 & 1.04 $\pm$ 1.91 & 425\\
1152.01 & 4.7222500 $\pm$ 8.6e-06 & 0.269 $\pm$ 3.0e-03 & $3806^{+50}_{-64}$ & -0.13 $\pm$ 0.10 & M1V & 0.52 $\pm$ 0.06 & 0.49 $\pm$ 0.06 & 0.04421 $\pm$ 1.64e-03 & 14.45 $\pm$ 1.69 & 560\\
1164.01 & 2.8011279 $\pm$ 8.8e-05 & 0.015 $\pm$ 5.5e-04 & $3684^{+55}_{-62}$ & 0.07 $\pm$ 0.10 & M1V & 0.51 $\pm$ 0.06 & 0.48 $\pm$ 0.06 & 0.03102 $\pm$ 1.20e-03 & 0.79 $\pm$ 0.10 & 642\\
1176.01 & 1.9737610 $\pm$ 1.0e-06 & 0.157 $\pm$ 2.9e-04 & $3806^{+388}_{-132}$ & -0.06 $\pm$ 0.19 & M1V & 0.53 $\pm$ 0.11 & 0.50 $\pm$ 0.11 & 0.02491 $\pm$ 1.66e-03 & 8.64 $\pm$ 1.82 & 756\\
1201.01 & 2.7575290 $\pm$ 4.6e-05 & 0.026 $\pm$ 2.8e-03 & $3712^{+85}_{-54}$ & -0.11 $\pm$ 0.11 & M1V & 0.48 $\pm$ 0.06 & 0.46 $\pm$ 0.06 & 0.03013 $\pm$ 1.21e-03 & 1.30 $\pm$ 0.22 & 638\\
1202.01 & 0.9283080 $\pm$ 1.5e-05 & 0.027 $\pm$ 8.5e-04 & $4894^{+607}_{-904}$ & -0.06 $\pm$ 0.12 & M0V & 0.76 $\pm$ 0.12 & 0.72 $\pm$ 0.11 & 0.01703 $\pm$ 8.80e-04 & 2.12 $\pm$ 0.34 & 1403\\
1266.01 & 11.4194403 $\pm$ 2.0e-04 & 0.027 $\pm$ 6.2e-04 & $3822^{+212}_{-99}$ & -0.21 $\pm$ 0.12 & M1V & 0.51 $\pm$ 0.08 & 0.49 $\pm$ 0.08 & 0.07932 $\pm$ 4.08e-03 & 1.45 $\pm$ 0.24 & 419\\
1298.01 & 11.0082302 $\pm$ 1.3e-04 & 0.044 $\pm$ 7.9e-03 & $5189^{+314}_{-798}$ & -0.37 $\pm$ 0.14 & K7V & 0.76 $\pm$ 0.12 & 0.72 $\pm$ 0.11 & 0.08852 $\pm$ 4.28e-03 & 3.44 $\pm$ 0.80 & 651\\
1361.01 & 59.8791008 $\pm$ 1.1e-03 & 0.034 $\pm$ 7.2e-04 & $3929^{+66}_{-135}$ & -0.02 $\pm$ 0.11 & M0V & 0.57 $\pm$ 0.07 & 0.55 $\pm$ 0.07 & 0.24861 $\pm$ 9.64e-03 & 2.04 $\pm$ 0.25 & 258\\
1403.01 & 18.7540894 $\pm$ 5.1e-04 & 0.028 $\pm$ 5.7e-02 & $3977^{+105}_{-184}$ & -0.20 $\pm$ 0.12 & M0V & 0.56 $\pm$ 0.08 & 0.54 $\pm$ 0.08 & 0.11380 $\pm$ 5.20e-03 & 1.65 $\pm$ 3.38 & 382\\
1404.01 & 6.6620498 $\pm$ 2.6e-04 & 0.024 $\pm$ 3.6e-02 & $3750^{+315}_{-105}$ & -0.23 $\pm$ 0.15 & M1V & 0.48 $\pm$ 0.10 & 0.45 $\pm$ 0.09 & 0.05409 $\pm$ 3.40e-03 & 1.19 $\pm$ 1.80 & 479\\
1408.01 & 14.5345097 $\pm$ 3.9e-04 & 0.022 $\pm$ 9.5e-04 & $4023^{+82}_{-113}$ & -0.11 $\pm$ 0.10 & M0V & 0.59 $\pm$ 0.07 & 0.57 $\pm$ 0.07 & 0.09751 $\pm$ 3.63e-03 & 1.36 $\pm$ 0.17 & 427\\
1422.01 & 5.8416181 $\pm$ 7.1e-05 & 0.038 $\pm$ 4.1e-02 & $3517^{+70}_{-57}$ & -0.07 $\pm$ 0.11 & M2V & 0.38 $\pm$ 0.06 & 0.37 $\pm$ 0.06 & 0.04595 $\pm$ 2.34e-03 & 1.53 $\pm$ 1.67 & 439\\
1422.02 & 19.8503704 $\pm$ 3.8e-04 & 0.036 $\pm$ 1.3e-03 & ... & ... & ... & ... & ... & 0.10386 $\pm$ 5.29e-03 & 1.45 $\pm$ 0.23 & 292\\
1422.03 & 3.6213870 $\pm$ 9.3e-05 & 0.025 $\pm$ 1.6e-03 & ... & ... & ... & ... & ... & 0.03341 $\pm$ 1.70e-03 & 1.01 $\pm$ 0.17 & 515\\
1427.01 & 2.6130109 $\pm$ 4.1e-05 & 0.021 $\pm$ 9.8e-04 & $3878^{+173}_{-75}$ & -0.20 $\pm$ 0.11 & M0V & 0.53 $\pm$ 0.07 & 0.51 $\pm$ 0.07 & 0.03003 $\pm$ 1.30e-03 & 1.16 $\pm$ 0.17 & 702\\
1459.01 & 0.6920230 $\pm$ 1.3e-06 & 0.075 $\pm$ 1.0e-03 & $3746^{+85}_{-84}$ & 0.07 $\pm$ 0.12 & M1V & 0.51 $\pm$ 0.07 & 0.49 $\pm$ 0.06 & 0.01222 $\pm$ 5.00e-04 & 3.98 $\pm$ 0.51 & 1043\\
1475.01 & 1.6093230 $\pm$ 2.4e-05 & 0.026 $\pm$ 1.3e-03 & $4056^{+417}_{-234}$ & -0.23 $\pm$ 0.13 & M0V & 0.57 $\pm$ 0.10 & 0.56 $\pm$ 0.09 & 0.02234 $\pm$ 1.18e-03 & 1.58 $\pm$ 0.27 & 894\\
1475.02 & 9.5124798 $\pm$ 2.5e-04 & 0.032 $\pm$ 1.5e-03 & ... & ... & ... & ... & ... & 0.07302 $\pm$ 3.84e-03 & 1.95 $\pm$ 0.33 & 494\\
1515.01 & 1.9370290 $\pm$ 2.1e-05 & 0.018 $\pm$ 5.6e-04 & $3944^{+80}_{-60}$ & -0.36 $\pm$ 0.10 & M0V & 0.52 $\pm$ 0.06 & 0.51 $\pm$ 0.06 & 0.02452 $\pm$ 8.80e-04 & 0.99 $\pm$ 0.12 & 790\\
1577.01 & 2.8062129 $\pm$ 6.0e-05 & 0.022 $\pm$ 1.5e-03 & $3940^{+161}_{-118}$ & -0.18 $\pm$ 0.12 & M0V & 0.55 $\pm$ 0.08 & 0.53 $\pm$ 0.08 & 0.03191 $\pm$ 1.47e-03 & 1.28 $\pm$ 0.20 & 710\\
1584.01 & 5.8708401 $\pm$ 1.5e-04 & 0.025 $\pm$ 1.1e-03 & $4187^{+999}_{-255}$ & -0.32 $\pm$ 0.14 & M0V & 0.59 $\pm$ 0.11 & 0.57 $\pm$ 0.10 & 0.05345 $\pm$ 3.02e-03 & 1.56 $\pm$ 0.28 & 604\\
1588.01 & 3.5174849 $\pm$ 4.5e-05 & 0.000 $\pm$ 1.8e+00 & $3916^{+86}_{-68}$ & -0.24 $\pm$ 0.10 & M0V & 0.53 $\pm$ 0.06 & 0.51 $\pm$ 0.06 & 0.03670 $\pm$ 1.36e-03 & 0.00 $\pm$ 0.00 & 643\\
1596.01 & 5.9236002 $\pm$ 1.3e-04 & 0.021 $\pm$ 2.2e-02 & $3880^{+143}_{-103}$ & -0.28 $\pm$ 0.12 & M0V & 0.51 $\pm$ 0.07 & 0.49 $\pm$ 0.07 & 0.05132 $\pm$ 2.27e-03 & 1.12 $\pm$ 1.19 & 529\\
1596.02 & 105.3551025 $\pm$ 3.9e-03 & 0.032 $\pm$ 7.1e-02 & ... & ... & ... & ... & ... & 0.34966 $\pm$ 1.55e-02 & 1.71 $\pm$ 3.81 & 202\\
\enddata
\tablenotetext{a}{Metallicity measurements for stars earlier than M0 ($T_{\rm eff} \lesssim 4000$ $K$) represent an extrapolation of the Rojas et al (2012) M dwarf [M/H] calibration.}
\tablenotetext{b}{The planetary equilibrium temperatures ($T_{\rm eq,P}$) were calculated assuming an Earth-like albedo and re-radiation fraction for each planet-candidate.  We do not include uncertainty estimates for $T_{\rm eq,P}$ due to the ambiguity of those assumptions.}
\tablenotetext{c}{There is no model in the Dartmouth isochrones with the same [M/H] and $T_{\rm eff}$ as KOI 961, so this star is interpolated onto the 5-Gyr BCAH isochrones.  See Muirhead et al. (2012) for a more detailed analysis of this star.}
\label{thetable}
\end{deluxetable}
\clearpage
\end{landscape}

\end{document}